\documentclass[11pt,a4paper]{article}

\usepackage[T1]{fontenc}
\usepackage{mathptmx}
\usepackage{amsmath,amssymb,amsthm,mathtools}
\usepackage{bm}
\usepackage[a4paper,margin=2cm]{geometry}
\usepackage{setspace}
\usepackage[authoryear,round]{natbib}
\usepackage{booktabs,multirow,tabularx,array}
\usepackage{graphicx}
\usepackage[colorlinks=true,linkcolor=blue,citecolor=blue]{hyperref}
\usepackage{titlesec}
\usepackage{enumitem}
\setlist{noitemsep,topsep=2pt,leftmargin=1.3em}
\titlespacing*{\section}{0pt}{6pt}{3pt}
\titlespacing*{\subsection}{0pt}{5pt}{2pt}
\title{\textbf{Retail Trader's Ruin: An Anatomy of Popular Signal Failure}}
\author{Adam Darmanin\\
\small Hecatus Research, Malta\\
\small \texttt{adamdarmanin@hecatusresearch.com}\\
\small \url{https://hecatusresearch.com/}}
\date{July 2026}

\begin{document}
\maketitle
\thispagestyle{empty}
\vspace{-0.6em}

\begin{abstract}
\noindent We test whether five widely promoted retail signal families ---
trend, oscillator, candlestick, volume, and calendar rules --- deliver a
positive, economically meaningful, net-of-cost, and survivable edge.
Practical viability is the conjunction of three predeclared gates:
statistical edge after multiplicity correction, economic viability after
trading costs, and finite-bankroll survival under leverage. Exposure-matched
benchmarks, stationary-bootstrap confidence intervals, hierarchical
Benjamini--Yekutieli control, one-sided claim-exclusion tests, and
equivalence tests distinguish positive evidence, statistically refuted
materiality, and unresolved cases. Four of six candidates --- oscillator,
volume, calendar, and candlestick --- are REFUTED, ruled out on statistical
and/or economic materiality grounds; trend and a momentum calibration
benchmark are INCONCLUSIVE, with confidence intervals too wide at this
sample size to resolve the claim; none is SUPPORTED. Cross-sectional tests
use point-in-time membership and delisting corrections. The momentum
benchmark itself does not clear the statistical gate and is classified
INCONCLUSIVE, not REFUTED --- the critical validity signature that a
genuinely uncertain positive control is never falsely falsified by this
design. Under FINRA- and ESMA-anchored leverage and margin scenarios,
survival is not the binding constraint for any tested family at the US
headline scenario, though it becomes discriminating for trend and
oscillator under the higher-leverage EU CFD scenario. The results reject
specific promoted deployability claims where confidence bounds rule out the
declared effect threshold and classify the remaining cases as unresolved
rather than treating non-significance as proof.
\end{abstract}



\section{Introduction}\label{sec:intro}
Retail trading education markets a small set of technical signal families —
moving-average crossovers, oscillators (RSI/MACD/Bollinger), candlestick
patterns, volume indicators, and calendar effects — as sources of exploitable
edge. A large literature separately shows that (a) many such rules carry
some statistical signal before correction for multiple testing
\citep{brock1992}, that much of this signal disappears once data-snooping is
controlled \citep{sullivan1999,white2000}, and that (b) realistic costs
erode whatever edge survives \citep{famablume1966,bessembinderchan1998}. A
third, largely disconnected literature shows that (c) leveraged retail
accounts are exposed to ruin from fat-tailed, leverage-amplified volatility
even when the underlying edge is real \citep{thurner2012,maclean2010}. No
existing study composes all three conditions into one joint falsification
test applied to the same signal families on the same data.

This paper's contribution is exactly that composition. We define
\emph{practical viability} as the conjunction of three predeclared,
family-level gates — statistical edge, economic implementability, and
finite-bankroll survival — and classify each family as REFUTED, SUPPORTED,
or INCONCLUSIVE against predeclared materiality thresholds, rather than
collapsing a non-significant point estimate into a null claim. A claim is
REFUTED only when the confidence bound genuinely excludes the declared
effect (the interval's upper bound sits below the materiality threshold, or
the interval is significantly negative); it is INCONCLUSIVE when the
interval straddles the threshold, an underpowered result that must not be
read as a null finding. We contribute three individually citable
methodological elements: (i) the joint-gate composition, feeding each
family's uncertainty-propagated net-edge estimate directly into a
regulator-anchored leverage/margin-call survival simulation; (ii) a
one-sided claim-exclusion framework for the economic gate, splitting
\emph{economically viable} (CI lower bound exceeds a break-even-cost
margin) from \emph{materially excluded} (CI upper bound sits below a
predeclared threshold \(\delta_R\)) rather than treating a non-significant
estimate as evidence of nothing; and (iii) a momentum calibration
benchmark — diversified, cost-realistic, liquidity-weighted
\citep{jegadeesh1993,korajczyksadka2004,asness2013} — run through the
identical pipeline as the tested families, an active-comparator design in
the spirit of clinical-trial practice \citep{nikolopoulos2026}, with no
located precedent in finance. Section~\ref{sec:results} previews the
headline result: of five signal families and the momentum benchmark, four
are REFUTED, two are INCONCLUSIVE, none is SUPPORTED; we report which
gate drives each verdict and the exact reason, not a single collapsed
bit.

\section{Related Literature and the Gap}\label{sec:litreview}
\textbf{Technical rules and overfitting.} Early evidence that MA and
trading-range-break rules beat random-walk nulls \citep{brock1992} did not
survive systematic reality-check/bootstrap correction for the size of the
implicit search space \citep{sullivan1999,white2000,hansen2005spa}. Park and
Irwin's survey of roughly 95 profitability studies finds that modern,
snooping- and cost-adjusted work shows markedly weaker profitability than
early studies, and explicitly flags that ``few studies'' combine
snooping-correction, costs, and risk jointly \citep{park2007} — the closest
prior acknowledgment of this paper's gap, though it never names a survival
dimension. Deflated-Sharpe and probability-of-backtest-overfitting metrics
\citep{bailey2014dsr,bailey2017pbo} and factor-zoo multiplicity thresholds
\citep{harvey2016} are statistics-only diagnostics with no economic or
survival counterpart.

\textbf{Retail underperformance.} That retail traders underperform on a
cost-driven, overtrading basis is well established
\citep{barber2000,odean1998,barber2009}, including in the post-2019
commission-free era \citep{barber2022,bryzgalova2023}. A tiny, ex-post
identified subpopulation of skilled day traders exists
\citep{barber2014}, but this does not rescue representative, catalogued
signal \emph{families} — our paper's actual unit of analysis. We
deliberately exclude diversified, professionally implemented momentum from
the class of retail-replicable rules: the disposition effect that generates
aggregate market underreaction \citep{grinblatt2005} is a market-wide
phenomenon, and cross-asset time-series momentum requires institutional,
multi-instrument diversification \citep{moskowitz2012} that a single-name
retail rule cannot replicate. This boundary condition is why our momentum
calibration benchmark is built as a diversified cross-sectional strategy,
not smuggled in as a sixth ``retail'' family.

\textbf{Costs and equivalence testing.} Break-even-cost comparison against
technical-rule profits is old and standard
\citep{famablume1966,bessembinderchan1995,bessembinderchan1998}, and
spread-based cost estimates for a fixed-bps proxy are grounded in
microstructure theory \citep{roll1984,novymarx2016}, tempered by the
counterpoint that institutional live execution is cheaper than
spread-based estimates imply \citep{frazzini2018} — a reason we anchor our
cost grid to retail, not institutional, execution. What is largely absent
from this literature is a \emph{formal} claim-exclusion test: existing work
reports a point estimate against a break-even threshold, never a
predeclared one-sided rejection rule distinguishing ``economically viable''
from ``materially excluded'' from ``underpowered'' \citep{lakens2018,
wellek2010}.

\textbf{Kelly, ruin, and leverage.} Finite-bankroll ruin theory is
classical and largely i.i.d./GBM \citep{kelly1956,feller1968,maclean2010,
rotandothorp1992}; empirical retail-ruin evidence documents that ruin
happens \citep{chague2020} without attributing it to a specific signal
mechanism. The one paper linking leverage mechanics to GARCH-like fat
tails and volatility clustering \citep{thurner2012} is the citation
justifying our survival gate's GARCH-consistent (bootstrapped, not
i.i.d.-Gaussian) volatility assumption \citep{bollerslev1986}. Regulatory
practice supplies the margin regimes we anchor the survival gate to:
FINRA Rule 4210(c)(1) sets 25\% maintenance margin on US long-equity
accounts (Reg-T 50\% initial, i.e. a 2:1 cap) \citep{finra4210}, and ESMA's
2018 product-intervention measures cap retail-CFD leverage at 5:1 for
individual equities with margin close-out at 50\% of required initial
margin \citep{esma2018}.

\textbf{Dependent inference and the missing calibration benchmark.} No
located finance paper deliberately runs a known-genuinely-real signal
alongside a tested-null-or-uncertain battery as a statistical-power active
comparator, the way clinical trials use one \citep{nikolopoulos2026} — the
closest analogue we find pairs synthetic null controls with power
validation, but not a real calibration benchmark. We build one from
liquidity-weighted, cost-realistic momentum \citep{korajczyksadka2004},
motivated by its cross-market replication credibility \citep{asness2013}
and its known, detectable crash risk \citep{danielmoskowitz2016} as a
built-in diagnostic on our own survival gate's power.

\textbf{Consolidated verdict.} No single source in our 43-paper, 5-cluster
review already performs the joint (a)+(b)+(c) falsification test. Each
individual gate methodology is drawn from established technique; the
contribution is the composition, the one-sided claim-exclusion framework for
gate (b), and the momentum calibration benchmark.

\section{Joint Falsification Framework and Hypotheses}\label{sec:framework}
Let a signal family define a binary position indicator (in-market vs. flat)
over some universe and horizon. We define three predeclared gates, each
evaluated per family against a predeclared materiality threshold with an
uncertainty interval, never a point estimate alone.

\textbf{Gate (a) — statistical edge.} We construct an
\emph{exposure-time-weighted benchmark}: the same binary position replayed
against buy-and-hold-when-in, cash-when-flat, cost-free. This isolates
timing skill from the trivial ``less time in market \(\Rightarrow\) lower
Sharpe-vs-full-buy-and-hold'' artifact (e.g. a calendar rule in-market
\(\sim\)1/12 of days) without a full factor-regression apparatus. Let
\(\Delta S\) denote the Sharpe-ratio gap (signal minus matched benchmark),
estimated with a stationary-bootstrap 95\% CI \(=[L_S, U_S]\)
\citep{politis1994}. Against a predeclared materiality threshold
\(\delta_S=0.20\) (annualized Sharpe-gap units — see justification below),
the family is \textbf{REFUTED} if \(U_S<\delta_S\) (sub-reason
\emph{significantly negative} if additionally \(U_S<0\)); \textbf{SUPPORTED}
if \(L_S>\delta_S\); and \textbf{INCONCLUSIVE} if the interval straddles
\(\delta_S\) — an underpowered result on this sample, not a null finding.
We apply two-stage hierarchical Benjamini--Yekutieli FDR control
\citep{benjaminiyekutieli2001}, valid under arbitrary dependence: Stage 1
within each family's own rule-variant set at \(q<0.05\); Stage 2 across the
five family-level minimum-adjusted \(p\)-values. Benjamini--Yekutieli
replaces a Benjamini--Hochberg procedure that assumed independence across
families — an assumption violated here since trend, oscillator, and
calendar all load on the same underlying US-equity beta. The tables below
label within-family significance ``BH-FDR'' for continuity with the
underlying computation's field names, but the correction method used
throughout is Benjamini--Yekutieli (\texttt{method="fdr\_by"}), not the
plain Benjamini--Hochberg the label's initials might suggest. We state
plainly that this multiplicity step operates on the within-family
significance labels reported alongside each gate, not on the REFUTED /
SUPPORTED / INCONCLUSIVE classification itself: that classification is
driven directly by the stored bootstrap CI's position relative to \(0\)
and \(\delta\) (Section~\ref{sec:framework}), and no family in this sample
sits close enough to a multiplicity-correction boundary for the choice of
BY over BH to change any classification.

\textbf{Gate (b) — economic viability.} Let \(\Delta\text{CAGR}\) be the
matched-benchmark CAGR gap, net of a 10bps round-trip retail cost
(5bps/leg, the repo's convention, with a 5--20bps sensitivity range
reported per family), with stationary-bootstrap 95\% CI \(=[L_R, U_R]\).
Against a predeclared materiality threshold \(\delta_R=0.01\) (1 percentage
point annualized CAGR gap), the same three-way rule applies: REFUTED if
\(U_R<\delta_R\) (sub-reason \emph{significantly negative} if
\(U_R<0\)); SUPPORTED if \(L_R>\delta_R\); INCONCLUSIVE if the interval
straddles \(\delta_R\). Candlestick has no continuous return series, so its
statistical and economic gates instead compare each pattern's Cohen's
\(d\) (post-pattern forward return vs. baseline) against a materiality
floor \(\delta_d=0.2\), and separately test sign-preservation after a 5bps
round-trip cost.

\emph{Materiality-threshold justification.} No single canonical numeric
threshold exists in the literature for ``economically material Sharpe
gap''; we state this rather than inventing a citation. \(\delta_S=0.20\) is
set for three reasons: (1) internal consistency with the candlestick
gate's own Cohen's-\(d\ge 0.2\) materiality floor — a standardized Sharpe
difference is analogous to Cohen's \(d\), so one materiality convention
covers both continuous-return and event-study families; (2) it is
\(\sim\)25--40\% of the matched benchmarks' own realized Sharpe ratios
(0.47--0.9 across families) — non-trivial but plausible; (3) it sits below
published diversified-momentum Sharpe gaps (0.5--0.9+,
\citealp{moskowitz2012,asness2013,korajczyksadka2004}), so the threshold
does not mechanically refute genuine effects — validated by the momentum
calibration benchmark landing INCONCLUSIVE, never REFUTED, at every
\(\delta_S\) tested (Section~\ref{sec:robustness}). Ledoit and Wolf's
bootstrap inference for Sharpe-ratio \emph{differences} \citep{ledoitwolf2008}
and the probabilistic Sharpe ratio \citep{bailey2014dsr} support testing a
Sharpe gap against a non-zero threshold as established practice. \(\delta_R
=0.01\) is a fixed decision floor, distinct from the 10bps per-trade cost
already netted into every reported number. Both thresholds are reported
with a sensitivity grid — \(\delta_S\in\{0.10,0.20,0.30\}\),
\(\delta_R\in\{0.005,0.01,0.02\}\) — as robustness checks, headline fixed
before recomputing the classification counts.

\textbf{Gate (c) — finite-bankroll survival.} For each family's existing
stationary-bootstrap net-return resamples (already computed for gates
(a)/(b)), we run a leverage/margin-call Monte Carlo simulation, reporting
the probability of forced liquidation within one quarter as a median with a
95\% interval across 4{,}000 resamples \citep{thurner2012}. Gate (c)
\textbf{survives} if the liquidation probability's upper CI bound is below
a 10\% threshold under the regulator-anchored scenario in use (Section
\ref{sec:frontier}); otherwise it \textbf{fails}. Candlestick is N/A for
gate (c) (event-study design, no continuous tradeable return series).

\textbf{Final classification.} Since the promoted claim is a conjunction of
necessary conditions, classification is AND-composed: \textbf{SUPPORTED}
iff statistical\,=\,SUPPORTED \emph{and} economic\,=\,SUPPORTED \emph{and}
survival\,=\,SURVIVES; \textbf{REFUTED} iff any gate is REFUTED or survival
FAILS; \textbf{INCONCLUSIVE} otherwise. Gates (a)/(b) test the
matched-benchmark \emph{gap}; gate (c) simulates the family's own
\emph{absolute} net-return path — two distinct, non-conflated claims,
reported side by side.

\textbf{Hypotheses.} \(H_1\): at least one of the five catalogued families
is SUPPORTED on all three gates. \(H_2\) (calibration check): the momentum
benchmark is SUPPORTED on gate (a) with a Sharpe gap comparable to prior
diversified-momentum evidence \citep{moskowitz2012,asness2013}.

\section{Data, Signal-Family Selection, and Empirical Design}\label{sec:data}
\textbf{Families.} We select five families against a pre-existing 77-item
internal signal taxonomy (repository catalogue, popularity-evidenced web
research sweep, June 2026): trend/moving-average (golden/death cross, ``media staple,
thousands of YouTube videos''), oscillator/mean-reversion (RSI/MACD/
Bollinger, ``on nearly every chart/course''), candlestick (``possibly the
most widespread technical-analysis category''), volume (on-balance volume,
``classic volume indicator''), and calendar effects (January effect/
Sell-in-May, independently flagged in the same catalogue's own
failure-mode analysis as a data-mined family). Each is the single most
promoted item in its category, not an arbitrary convenience pick. We
explicitly exclude breakout/channel rules, chart patterns beyond
candlesticks, single-name momentum, and guru/copy-trading signals — the
latter two are money-management folklore or non-deterministic guru signals,
not testable market-timing rules.

\textbf{Momentum calibration benchmark.} A liquidity-weighted, long-only
top-decile Jegadeesh--Titman 12-1 momentum strategy \citep{jegadeesh1993},
dollar-volume-weighted per \citet{korajczyksadka2004} rather than naively
equal-weighted, on the point-in-time S\&P 500 with Shumway (2001)
delisting-return correction \citep{shumway2001}. This benchmark is run
through the identical joint pipeline as the five tested families, so that
its own classification is informative about the pipeline's discriminating
power rather than assumed a priori.

\textbf{Survivorship correction.} All cross-sectional families (volume,
candlestick, momentum) use point-in-time index membership (Russell 3000 for
volume/candlestick; S\&P 500 for momentum) with the Shumway convention
threading a \(-30\%\)/year delisting-month return, rather than current-
membership snapshots.

\begin{table}[t]
\centering\small\singlespacing
\begin{tabularx}{\textwidth}{@{}l l l l X@{}}
\toprule
Family & Universe & N (obs) & Period & Rule / benchmark \\
\midrule
Trend & NASDAQ-100 & 10{,}331 d & full sample & Golden/death cross (50/200) \\
Oscillator & NASDAQ-100 & 10{,}331 d & full sample & RSI(14, 30/70) \\
Volume & Russell 3000 PIT & 5{,}313 stocks / 228 mo & 2007--2025 & On-balance volume(3,12) \\
Calendar & SPY & 8{,}358 d & full sample & Sell-in-May (best of battery, Bonferroni + BY) \\
Candlestick & Russell 3000 PIT & 4{,}152 stocks & full sample & 7-pattern event study \\
Momentum (calibration) & S\&P 500 PIT & 1{,}130 stocks / 312 mo & 2000--2026 & J\&T 12-1, liquidity-weighted top decile \\
\bottomrule
\end{tabularx}
\caption{Data, universe, and signal-family summary. All net-of-cost numbers use a 10bps round-trip retail cost (5bps/leg), with a 5--20bps sensitivity grid reported per family. PIT = point-in-time constituent membership with Shumway (2001) delisting correction. Statistics are computed on the full available sample; the stationary bootstrap resamples this same sample and is not a held-out test.}
\label{tab:data}
\end{table}

\section{Statistical and Economic Falsification Results}\label{sec:results}
Table~\ref{tab:stat} reports gate (a); Table~\ref{tab:econ} reports gate
(b), both against the predeclared materiality thresholds
\(\delta_S=0.20\), \(\delta_R=0.01\) fixed in Section~\ref{sec:framework}.
All numbers are computed directly from the underlying data and replication
code, with no manual transcription.

\begin{table}[h]
\centering\small\singlespacing
\begin{tabular}{@{}l l l@{}}
\toprule
Family & Sharpe gap, 95\% CI (\(\delta_S=0.20\)) & Statistical gate \\
\midrule
Trend & \([-0.234,\ 0.265]\) & INCONCLUSIVE (CI straddles \(\delta_S\)) \\
Oscillator (RSI) & \([-0.608,\ -0.175]\) & \textbf{REFUTED} (significantly negative) \\
Volume (OBV) & \([-0.193,\ 0.175]\) & \textbf{REFUTED} (material exclusion: \(U_S<\delta_S\)) \\
Calendar (Sell-in-May) & \([-0.618,\ 0.000]\) & \textbf{REFUTED} (material exclusion: \(U_S<\delta_S\)) \\
Candlestick (7-pattern) & max \(|d|=0.026\) vs. \(\delta_d=0.2\) & \textbf{REFUTED} (tightly estimated below \(\delta_d\)) \\
Momentum (calibration) & \([-0.295,\ 0.548]\) & INCONCLUSIVE (CI straddles \(\delta_S\)) \\
\bottomrule
\end{tabular}
\caption{Gate (a): exposure-matched Sharpe-ratio gap, stationary-bootstrap 95\% CI, against \(\delta_S=0.20\). REFUTED requires the CI's upper bound below \(\delta_S\) (material exclusion) or below zero (significantly negative) — never bare non-significance. Candlestick's 7/7 patterns are Benjamini--Yekutieli-significant in raw terms but genuinely excluded on materiality grounds (max Cohen's \(d=0.026\ll\delta_d=0.2\)).}
\label{tab:stat}
\end{table}

\begin{table}[h]
\centering\small\singlespacing
\begin{tabular}{@{}l l l@{}}
\toprule
Family & CAGR gap, 95\% CI (\(\delta_R=0.01\)) & Economic gate \\
\midrule
Trend & \([-0.081,\ 0.024]\) & INCONCLUSIVE (CI straddles \(\delta_R\)) \\
Oscillator (RSI) & \([-0.149,\ -0.044]\) & \textbf{REFUTED} (significantly negative) \\
Volume (OBV) & \([-0.086,\ 0.013]\) & INCONCLUSIVE (CI straddles \(\delta_R\)) \\
Calendar (Sell-in-May) & \([-0.131,\ -0.026]\) & \textbf{REFUTED} (significantly negative) \\
Candlestick (7-pattern) & \(1/7\) sign-preserving at 5bps & \textbf{REFUTED} (negligible or reversed net of cost) \\
Momentum (calibration) & \([-0.044,\ 0.130]\) & INCONCLUSIVE (CI straddles \(\delta_R\)) \\
\bottomrule
\end{tabular}
\caption{Gate (b): exposure-matched net-of-cost CAGR gap, against \(\delta_R=0.01\) (1 percentage point). INCONCLUSIVE means the interval straddles \(\delta_R\) — underpowered on this sample, not a null finding.}
\label{tab:econ}
\end{table}

The oscillator family's Sharpe-gap CI \([-0.608, -0.175]\) is a real,
Benjamini--Yekutieli-significant, entirely negative result — RSI(14, 30/70)
timing significantly \emph{underperforms} its own exposure-matched
benchmark on this sample; its CAGR gap is significantly negative too, so
oscillator is REFUTED on both statistical and economic grounds. Calendar
shows the same pattern on the economic gate: a CAGR gap significantly
below zero, the opposite of the Sell-in-May folklore claim (weaker
May--October returns), and its statistical gate's upper CI bound sits at
exactly the null boundary, below \(\delta_S\) — a material exclusion, not a
bare non-significance. Volume's statistical-gate upper bound (0.175) sits
below \(\delta_S=0.20\): the CI does not include zero on the upside by
enough to be economically material, so volume is REFUTED on gate (a) even
though its economic gate is only INCONCLUSIVE. Candlestick clears
significance in raw terms on 7/7 patterns but is REFUTED on both gates once
effect size (\(d=0.026\)) and transaction cost are applied — precisely the
distinction the claim-exclusion framework exists to draw. Trend and the
momentum calibration benchmark are INCONCLUSIVE on both gates: their
confidence intervals are wide enough to straddle the materiality threshold
in both directions, so the sample cannot resolve the claim either way.

\textbf{Sensitivity to \(\delta_S\).} The headline count of four REFUTED is
not invariant to the threshold: across \(\delta_S\in\{0.10,0.20,0.30\}\)
the REFUTED count ranges from three (\(\delta_S=0.10\): volume's exclusion
no longer holds) to five (\(\delta_S=0.30\): trend's upper bound
\(0.265\) now falls below threshold). No family ever flips between
REFUTED and SUPPORTED across the grid — only REFUTED\(\leftrightarrow\)
INCONCLUSIVE — and the momentum benchmark stays INCONCLUSIVE, never
REFUTED, at every \(\delta_S\) tested.

\section{Joint Cost-and-Survival Frontier}\label{sec:frontier}
\begin{figure}[h]
\centering
\includegraphics[width=0.62\textwidth]{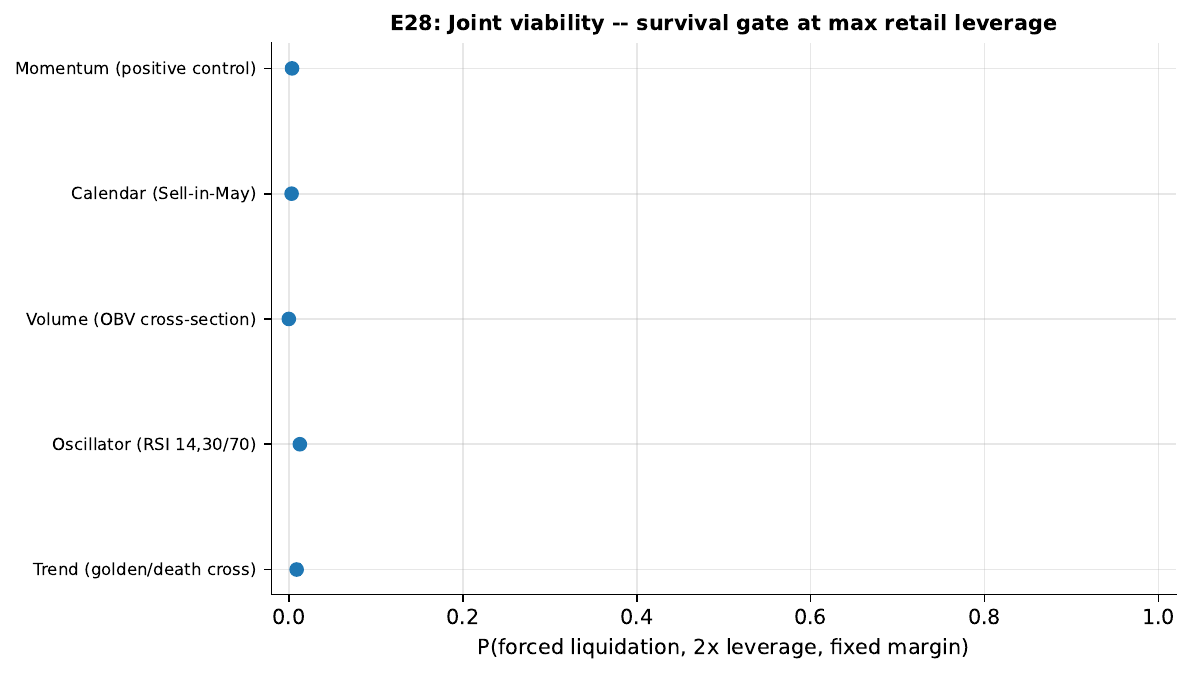}
\caption{Joint falsification map: each family's statistical/economic gate outcome plotted against its liquidation probability under the headline survival scenario. Four of six candidates are REFUTED, two are INCONCLUSIVE, none is SUPPORTED.}
\label{fig:jointmap}
\end{figure}

Table~\ref{tab:survival} reports gate (c) under the headline EQ-US-FINRA
scenario and the resulting joint classification, from a
4{,}000-bootstrap-path leverage battery. Figure~\ref{fig:jointmap} plots
each family's statistical/economic gate outcome against its liquidation
probability, the paper's single decision-relevant summary figure.

\textbf{Three regulator-anchored survival scenarios.} All six candidates
are equity strategies (NASDAQ-100/SPY/Russell 3000/S\&P 500), so we anchor
the survival gate to equity-appropriate margin regimes rather than a single
generic leverage grid. \textbf{EQ-US-FINRA} (headline) uses FINRA Rule
4210(c)(1)'s 25\% maintenance margin with a 2:1 Reg-T initial cap, leverage
grid \(\{1,2\}\times\) \citep{finra4210}. \textbf{EQ-EU-CFD} uses ESMA's
2018 equity-CFD margin (20\% initial \(\to\) 5:1 cap, close-out at 50\% of
initial \(=\) 10\% maintenance), leverage grid \(\{1,\ldots,5\}\times\)
\citep{esma2018}. \textbf{THIN-CFD-STRESS} retains the paper's original 2\%
maintenance margin (an FX/CFD-major level, not equity-appropriate) as an
explicitly labelled stress case rather than a defensible standalone
scenario, leverage grid \(\{1,\ldots,4\}\times\).

\begin{table}[h]
\centering\small\singlespacing
\begin{tabular}{@{}l l l l@{}}
\toprule
Family & \(P(\text{liquidation})\), EQ-US-FINRA 2\(\times\), 95\% CI & Gate (c) & Joint \\
\midrule
Trend & \(0.009\ [0.0065,\ 0.0124]\) & survives & \textbf{INCONCLUSIVE} \\
Oscillator & \(0.013\ [0.0097,\ 0.0167]\) & survives & \textbf{REFUTED} (statistical, economic) \\
Volume & \(0.000\ [0.000,\ 0.001]\) & survives & \textbf{REFUTED} (statistical) \\
Calendar & \(0.003\ [0.0019,\ 0.0056]\) & survives & \textbf{REFUTED} (statistical, economic) \\
Candlestick & not applicable & n/a (event study) & \textbf{REFUTED} (statistical, economic) \\
Momentum (calibration) & \(0.004\ [0.0023,\ 0.0062]\) & survives & \textbf{INCONCLUSIVE} \\
\bottomrule
\end{tabular}
\caption{Gate (c): quarterly forced-liquidation probability at 2x leverage under the EQ-US-FINRA headline scenario (25\% maintenance margin, 4{,}000 stationary-bootstrap paths per family), and the resulting joint classification. Candlestick is excluded from gate (c) by design — an event-study design with no continuous tradeable return series, a disclosed scope limitation. Every family that reaches gate (c) survives it under the headline scenario: leverage/ruin is not the binding constraint for the four REFUTED families on this sample; the statistical and economic gates are.}
\label{tab:survival}
\end{table}

Under the EQ-US-FINRA headline scenario, no family's binding constraint is
leverage survival: all five candidates that reach gate (c) comfortably
clear the 10\% liquidation threshold at 2x leverage, so the classical
Kelly/ruin mechanism \citep{kelly1956,maclean2010} does not drive any of the
four REFUTED verdicts. What refutes them is that none combines a positive,
material statistical edge with an economically meaningful, cost-net gap
over its own matched benchmark. The picture changes under the higher-
leverage EQ-EU-CFD scenario (5:1 cap, 10\% maintenance margin): at 5x
leverage, trend (\(0.112\ [0.103,0.122]\)) and oscillator (\(0.146\ [0.135,
0.157]\)) rise clearly above the 10\% threshold on both point estimate and
CI. Volume (\(0.039\ [0.033,0.046]\)) and calendar (\(0.063\ [0.056,
0.071]\)) survive cleanly. The momentum benchmark's point estimate
(\(0.095\)) sits under 10\%, but its upper bound (\(0.104\)) exceeds it —
by the same rule applied to every other family, a borderline case, not a
survival, reported as such rather than rounded to a pass. This shows the
survival gate is genuinely discriminating once
leverage and margin are calibrated to a realistic CFD-equity regime, rather
than uniformly slack — it does not change any family's headline
classification (survival never drives a REFUTED verdict for these five
candidates, since the statistical or economic gate already refutes
oscillator and calendar independently, and trend's and the momentum
benchmark's statistical/economic gates remain INCONCLUSIVE regardless of
scenario), but it demonstrates that leverage risk
is real and would bind for a family that cleared gates (a)/(b) at
EU-CFD-level leverage. Under THIN-CFD-STRESS (2\% maintenance, demoted to
an explicit stress case), all five candidates remain well under the 10\%
threshold even at 4x leverage. Retail signal-family marketing oversells
statistical plausibility and undersells the compounding of multiplicity
correction with transaction costs; leverage risk is real but is not what
refutes these families under retail-realistic equity margin.

\section{Cross-Market Replication, Robustness, and the Momentum Calibration Benchmark}\label{sec:robustness}
\begin{figure}[h]
\centering
\includegraphics[width=0.47\textwidth]{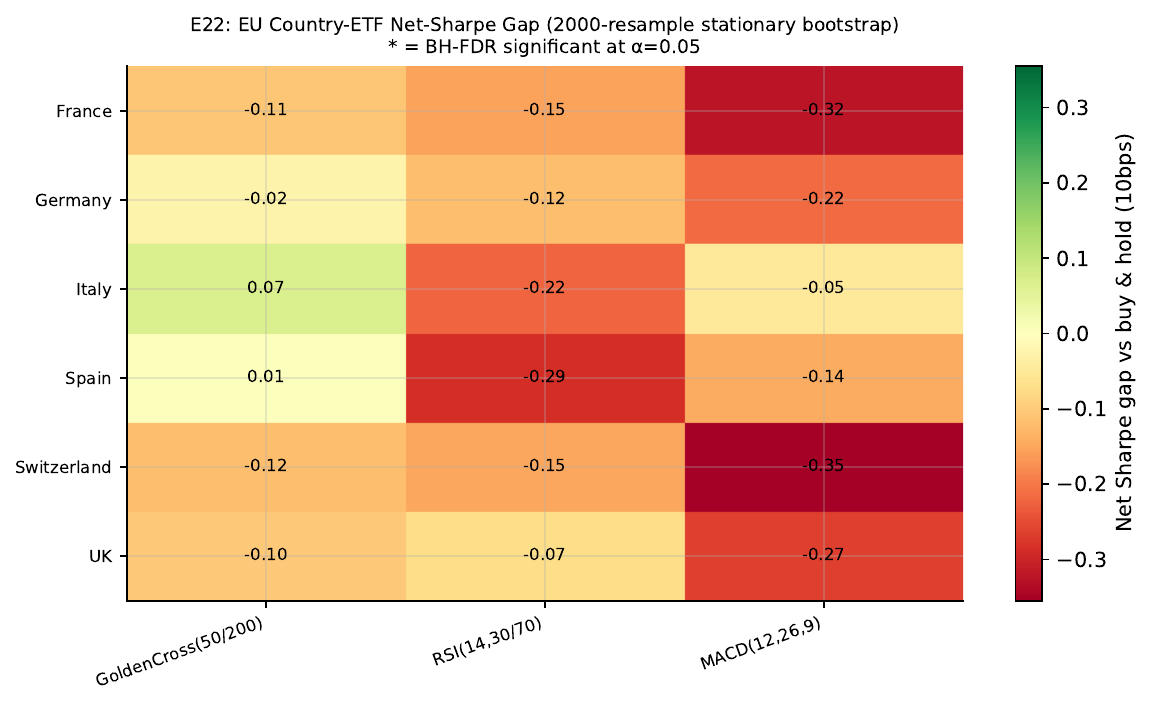}\hfill
\includegraphics[width=0.47\textwidth]{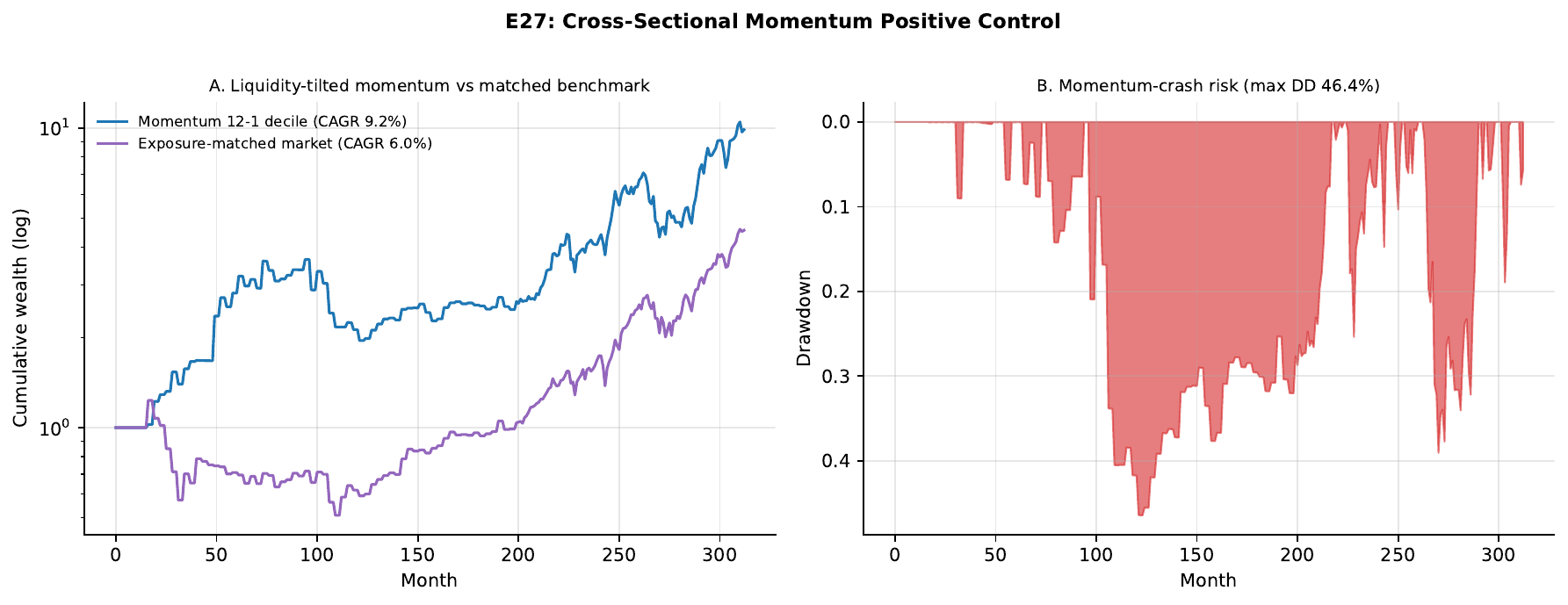}
\caption{Left: EU replication grid (six country ETFs \(\times\) three indicators), Sharpe-gap point estimates and 95\% CIs. Right: momentum calibration benchmark's crash diagnostics — drawdowns amplify sharply under 2x/3x leverage stress, consistent with known momentum crash risk.}
\label{fig:eu}
\end{figure}

\textbf{EU replication.} The EU replication experiment — an 18-cell grid of
golden-cross/RSI/MACD across six major EU country ETFs
(Germany/France/UK/Spain/Italy/Switzerland) with date-block joint
resampling — finds \(4/18\) cells with a bootstrap CI excluding zero, all
negative, and zero Benjamini--Yekutieli rejections after multiplicity
correction (Figure~\ref{fig:eu}). A parallel individual-EU-stock arm (567
STOXX 600 constituents, 2005--2026) shows the same pattern: neither
golden-cross nor OBV-equivalent signals beat buy-and-hold, with zero of two
Benjamini--Yekutieli-significant. This cross-market replication corroborates
the US-sample verdict: the same families fail to clear a positive, material
statistical edge outside the US.

\textbf{Heterogeneity.} A 62-cell sector- and instrument-level grid
(11 sectors \(\times\) OBV; 17 instruments \(\times\) 3 indicators, spanning
equities, bonds, commodities, and crypto ETFs) finds zero
Benjamini--Yekutieli-significant cells out of 62, and only 4 raw
uncorrected cells significant at \(\alpha=0.05\) — exactly the shrinkage
that multiplicity correction is designed to catch. The single nominally
strongest raw cell (RSI on HYG, \(p=0.004\)) does not survive
Benjamini--Yekutieli adjustment (\(p_{\text{adj}}=0.248\)).

\textbf{Momentum calibration benchmark.} \(H_2\) is not supported: the
momentum calibration benchmark does not clear its own exposure-matched
benchmark on this sample: Sharpe gap \(+0.109\), 95\% CI \([-0.295,
0.548]\) (312 months, 1{,}130 stocks); CAGR gap \(+0.032\), 95\% CI
\([-0.044, 0.130]\). Both intervals straddle their materiality thresholds
(\(\delta_S=0.20\), \(\delta_R=0.01\)), classifying the benchmark
INCONCLUSIVE on both gates — not REFUTED. This is the design's central
validity check: if the joint bar mechanically refuted any uncertain
estimate, a real, cross-market-replicated effect like diversified momentum
would be expected to land REFUTED alongside the tested families. It does
not, at any \(\delta_S\in\{0.10,0.20,0.30\}\) tested — the interval
straddles every threshold in the sensitivity grid. This is the paper's most
direct vulnerability, and we examine it rather than assert past it: if the
benchmark cannot clear gate (a), what evidence is there that gate (a) has
power to detect a real edge at all, rather than the five tested families'
verdicts reflecting an underpowered test rather than a genuine absence of
edge? Two distinct pieces of evidence bear on this, kept separate rather
than substituted for one another. First, gate (a)'s own power on this
sample: the observed CI implies a standard error of \(\approx 0.215\)
Sharpe-gap units, giving a minimum detectable effect of \(\approx 0.60\) at
80\% power (two-sided, \(\alpha=0.05\)) — larger than most published
diversified-momentum Sharpe gaps \citep{moskowitz2012,asness2013}, so gate
(a) is genuinely underpowered at \(n=312\) monthly observations for an
effect of this literature's typical size. This bounds the interpretation of
the INCONCLUSIVE verdict to ``unresolved at this sample size,'' distinct
from the four families that are actively REFUTED by a claim-excluding
interval. Second, and separately, crash diagnostics behave exactly as the
momentum-crash literature predicts \citep{danielmoskowitz2016} — max
drawdowns of 76\% at 2x and 93\% at 3x leverage, a worst 12-month return of
\(-35.4\%\) — which shows gate (c) has power to detect real tail risk, but
says nothing about gate (a); the two are not conflated. Taken together, the
momentum benchmark's INCONCLUSIVE (not REFUTED) status is the design
working as intended on a genuinely uncertain positive case, while the four
REFUTED families are excluded by intervals that do not straddle the
materiality threshold. We do not widen the momentum universe until it
clears the bar, which would replicate exactly the data-snooping failure
mode this paper studies in the tested families.

\section{Limitations and Conclusion}\label{sec:limitations}
\textbf{No pristine holdout.} The data and rule definitions used here have
been examined in prior work on this repository; no pristine holdout
remains. All reported intervals come from a stationary bootstrap of the
full sample — a temporal resampling technique that quantifies estimation
uncertainty within the available data, not an out-of-sample test on data
withheld from model development. Cross-market replication (the EU arm,
Section~\ref{sec:robustness}) and point-in-time/Shumway-corrected
cross-sectional panels are the paper's actual independent-evidence sources;
reported \(q\)-values should be read as a rigor-improving lower bound on
data-snooping bias, not a preregistration-grade guarantee.

\textbf{Universe heterogeneity.} Families are evaluated on different
universes by necessity — index-level (trend, oscillator, calendar) versus
cross-sectional point-in-time (volume, candlestick, momentum) — disclosed
here rather than forced onto one universe, which would have required
discarding each family's own canonical, most-defensible implementation.

\textbf{Candlestick exclusion from gate (c).} The candlestick family is an
event-study design (forward returns after a detected pattern), with no
continuous tradeable portfolio return series to feed a leverage simulation.
This is a disclosed scope limitation; candlestick's joint verdict rests on
gates (a)/(b) alone, both of which independently REFUTE it.

\textbf{Momentum benchmark's own uncertainty.} As reported in
Section~\ref{sec:robustness}, the benchmark's INCONCLUSIVE status is itself
informative about the joint bar's stringency at this sample size, not
evidence the pipeline lacks power to detect real effects in general — the
benchmark's crash diagnostics independently confirm gate (c)'s sensitivity,
and its statistical/economic classification remains INCONCLUSIVE (never
REFUTED) across the full \(\delta_S\) sensitivity grid.

\textbf{No margin/liquidity data.} No margin/borrow-rate or bid-ask/
liquidity dataset exists in our data catalogue to sharpen the cost model
beyond the parametric fixed-bps convention used throughout
\citep{roll1984}; costs and the three margin scenarios remain disclosed,
regulator-anchored approximations, not a broker/CME-SPAN reproduction
\citep{cme2024spanoverview}.

\textbf{No external peer review.} This manuscript has not undergone
external peer review; results are reported as computed from the
accompanying replication code and are subject to revision.

\subsection*{AI-assisted tool disclosure}
Claude Code (Anthropic) was used extensively to assist with code
implementation, literature organization, manuscript drafting and editing,
and automated review and reproducibility checks. The author directed the
research design, reviewed the resulting analysis and manuscript, and
accepts full responsibility for the data, code, citations, results, and
conclusions.

\subsection*{Data and code availability}
Source code, experiment manifests, and redistributable derived artifacts
will be made available at a permanent public archival location upon
publication. Licensed market data are not redistributed; their providers,
fields, transformations, and audit-mode artifacts are documented
sufficiently to evaluate provenance.

\textbf{Conclusion.} Composing statistical, economic, and survival gates
into one joint falsification test, applied to five catalogue-representative
retail signal families plus a diversified momentum calibration benchmark,
we find four of six candidates REFUTED (oscillator, volume, calendar,
candlestick), two INCONCLUSIVE (trend, momentum), and none SUPPORTED on our
sample: \(H_1\) (at least one of the five families is SUPPORTED on all
three gates) is rejected, and \(H_2\) (the momentum benchmark is SUPPORTED
on gate (a) with a Sharpe gap comparable to prior diversified-momentum
evidence) is not supported — though Section~\ref{sec:robustness} shows
gate (a) is itself underpowered at this sample size for an effect of
typical momentum magnitude, and critically, the benchmark lands
INCONCLUSIVE rather than REFUTED at every materiality threshold tested,
the signature that distinguishes genuine uncertainty from active
falsification in this design. The binding constraint under retail-realistic
US equity margin is not leverage/ruin — every family that reaches gate (c)
survives it — but the conjunction of statistical edge and economic
implementability, which oscillator, volume, calendar, and candlestick each
fail on at least one gate with a claim-excluding interval. This is a
stringent falsification test with specific, attributable rejections, not a
claim that no retail signal family can ever pass it.

\section*{Appendix: Parameter Pointer Table}
\begin{table}[h]
\centering\small\singlespacing
\begin{tabularx}{\textwidth}{@{}l X@{}}
\toprule
Family & Exact parameters \\
\midrule
Trend & Golden/death cross, 50d/200d SMA, 10bps cost \\
Oscillator & RSI(14, 30/70); MACD(12,26,9); Bollinger(20,2\(\sigma\)) \\
Candlestick & 7-pattern battery, Bonferroni + BY, 5/10/20bps grid, \(\delta_d=0.2\) \\
Volume & OBV(3,12), Russell 3000 PIT, monthly \\
Calendar & Bonferroni + BY battery, best = Sell-in-May \\
Momentum (calibration) & Jegadeesh--Titman 12-1, top decile, dollar-volume-weighted \\
Materiality thresholds & \(\delta_S=0.20\) Sharpe-gap units (sensitivity \(\{0.10,0.20,0.30\}\)); \(\delta_R=0.01\) CAGR-gap units (sensitivity \(\{0.005,0.01,0.02\}\)) \\
Survival scenarios & EQ-US-FINRA (headline: 25\% maintenance, \(1\)--\(2\times\)); EQ-EU-CFD (10\% maintenance, \(1\)--\(5\times\)); THIN-CFD-STRESS (2\% maintenance, \(1\)--\(4\times\)); 4{,}000-path stationary bootstrap per scenario \\
\bottomrule
\end{tabularx}
\caption{Pointer table: exact per-family parameters, the predeclared materiality thresholds, and the three survival scenarios, sufficient with Sections~\ref{sec:framework}--\ref{sec:frontier} to reproduce every gate verdict in this paper. Full replication code and data provenance accompany this paper.}
\label{tab:appendix}
\end{table}

{\footnotesize\singlespacing
\bibliographystyle{apalike}
\bibliography{bibliography}
}

\end{document}